\documentclass[aps,groupedaddress,showpacs]{revtex4}
\usepackage{graphicx}
\usepackage{amsmath}

\def\be{\begin{equation}}
\def\ee{\end{equation}}
\def\bea{\begin{eqnarray}}
\def\eea{\end{eqnarray}}

\begin{document}

\title{An Environmental Variation of Constants}

\author{Philippe Brax}
\affiliation{Institut de Physique Th\'eorique, CEA, IPhT, CNRS, URA 2306,
  F-91191Gif/Yvette Cedex, France}
\email{philippe.brax@cea.fr}

\date{\today}

\begin{abstract}
Models of modified gravity, whereby local tests of gravity are evaded thanks to a screening mechanism  of the
chameleon or Damour-Polyakov types, lead to a spatial variation of the particle masses and the fine structure constant.
This is triggered by the environmental dependence of the value of the  scalar field whose presence modifies gravity. In dense media, the field settles at a density dependent value
while in sparse environments it takes the background cosmological value. We estimate that the maximal deviation of constants from their present values is constrained by  local tests of gravity, and must be  less than $10^{-6}$.
\end{abstract}

\maketitle

\section{Introduction}

The acceleration of the expansion of the Universe may be due to the dynamics of a scalar degree of freedom whose energy density would lead to the existence of Dark Energy \cite{Astier:2012ba,Copeland:2006wr}.
It could also be that the acceleration occurs because the laws of gravity are modified on large scales and deviate from General Relativity. Recently, it has been realised that
large classes of models of modified gravity also involve a scalar field \cite{Khoury:2010xi}. All in all, scalar fields may be at the heart of unexpected phenomena on the largest scales of the Universe.

In both dark energy and modified gravity models, the scalar field acquires a mass which is extremely low and therefore acts as a nearly massless field in the solar system.
When coupled to matter, this would lead to large deviations from Newton's law which have not been  observed so far. This has motivated the construction of screening mechanisms \cite{KhouryWeltman,Khoury:2003aq,Brax:2004qh,Damour:1994zq,Pietroni:2005pv,Olive:2007aj,Hinterbichler:2010es,Brax:2010gi} whereby a scalar field can appear nearly massless on cosmological scales and still evade gravitational tests locally. One fundamental property of these models is that the scalar field settles at the minimum of the  effective potential inside screened bodies, a value which is matter density dependent.
For instance, the  scalar field would be different inside the sun and in the bulk of the Milky Way.

In such scalar-tensor theories, particle masses and, through quantum effects, the fine structure constant depend on the scalar field value \cite{Brax:2010uq}. In an environment where some dense regions of space are screened while sparser ones are unscreened, the scalar field develops a spatially dependent profile where its value  varies from a density dependent region to the cosmological one. Of course this would lead to a spatial variation of both particle masses and the fine structure constant. Such a  spatial dependence of fundamental constants was first noted in the context of the BSBM theory \cite{Sandvik:2001rv,Magueijo:2002di,Barrow:2002kz,Barrow:2002db,Mota:2003tm,Barrow:2011kr,Li:2010zt} where the variation of the fine structure constant is locally proportional to the variation of  Newton's potential. The resulting proportionality coefficient is strongly constrained by Earth-bound atomic clock experiments \cite{Leefer:2013waa}. A variation of both the fine structure constant and particle masses was also obtained in  dilatonic scalar-tensor theories with no scalar potential \cite{Flambaum:2007ar},  and shown to be proportional to the variation of Newton's potential too. Finally, the first study of the variation of constants in a screened modified gravity model was carried out in \cite{Olive:2007aj} where the scalar potential and the coupling function have two different minima where the field settles depending on the matter density. We will generalise the latter \cite{Silva:2013sla} to all screened models where the screening mechanism is of the chameleon or Damour-Polyakov types \footnote{The Olive-Pospelov model provides an explicit example of Damour-Polyakov screening.}.  In these theories, the measurements of the variations of both the proton to electron mass ratio and the fine structure constant does not show a direct dependence on the redshift of the absorbing system where atomic transitions take place. On the contrary, two absorbing systems located at the same distance, or the same redshift, could perfectly lead to unrelated particle masses and fine structure constant value if they happen to be in different environments characterised by different densities, all the more if one sits in a screened astrophysical object (such as a galaxy) while the other one lies in an unscreened region (such as an intergalactic Lyman limit system). In this paper, we will analyse the environment dependence of fundamental constants in these modified gravity models. We will apply bounds on the scalar field interaction range on cosmological scales to obtain new constraints on the environmental dependence of the proton to electron mass ratio and the fine structure constant.

In section II, we recall some of the properties of screened modified gravity, focusing on $f(R)$ and dilaton models. We also discuss the screening criterion and the local tests of gravity. In section III, we calculate the
variation of the proton to electron mass ratio $\mu=\frac{m_p}{m_e}$ and the fine structure constant $\alpha$ in these models. In section IV, we compare our results to the present observational bounds and give some prospects
on the possibility of observing screened modified gravity effects in the future. We then conclude.

\section{Screened Modified Gravity}

\subsection{Modified Gravity}

We focus on theories which modify gravity in the Einstein frame where the Einstein-Hilbert term is not altered
\be
S_{EH}=\int d^4x \sqrt{-g} \frac{R}{16\pi G_N},
\ee
 $R$ is the Ricci scalar of  the metric $g_{\mu\nu}$ and we have identified $8\pi G_N= m_{\rm Pl}^{-2}$ where $m_{\rm Pl}$ is the reduced Planck scale. This is supplemented with the scalar part of the action which involves both the scalar and matter fields. In general such a field theory is  very  complex and in a given environment involving macroscopic bodies comprising  non-relativistic matter the theory admits a vacuum configuration $\phi_0$ which depends on the distribution of matter.
At this level,
the scalar field couples to matter inhomogeneities  via the coupling constant $\beta(\phi_0)$.
 The mass of the scalar field $m(\phi_0)$ depends on the environment too.
Gravity is modified in as much as the coupling of $\phi$ to matter implies a modification of the geodesics compared to General Relativity. They  depend  on the full  potential
\be
\Phi =\Phi_N + \beta(\phi_0) \frac{\phi}{m_{\rm Pl}}
\ee
where $\Phi_N $ is the Newtonian potential satisfying the  Poisson equation.
The chameleon mechanism \cite{KhouryWeltman,Khoury:2003aq} occurs when the mass $m(\phi_0)$ is large enough to suppress the range of the scalar force in dense environments. The Damour-Polyakov  screening is such that $\beta(\phi_0)$ itself is small \cite{Damour:1994zq}.

More precisely, in screened objects the mass is so large or the coupling so small that the scalar field is essentially constant. This is enough to screen the effect of the scalar field outside a massive body.
Denoting by $\phi_c$ the value inside the object  and by $\phi_\infty$ the value outside and far away from the body, an approximate solution of the Klein-Gordon equation in the spherical case which describes accurately the outside solution in the screened case \cite{Brax:2012gr}, is simply
\be
\phi(r)=\phi_\infty  -2Qm_{\rm Pl}\frac{G_N M}{r}
\ee
where $M$ is the mass of the dense object and $R$ its radius. We have defined the scalar charge
\be
Q=\frac{\phi_\infty-\phi_c}{2m_{\rm Pl}\Phi_N}
\ee
where $\Phi_N$ is the value of Newton's potential at the surface of the body $\Phi_N= \frac{G_N M}{R}$.
The scalar charge depends on the environment via $\phi_\infty$ and  on the properties of the body via $\phi_c$ and $\Phi_N$.
Comparing to the linear solution for a point-like source,
we immediately recover that the screening criterion \cite{KhouryWeltman} for scalar-tensor theories is
\be
Q\lesssim  \beta_\infty
\ee
which requires that the scalar charge of a screened object should be  smaller than the coupling to matter far away from the object. This type of screening is entirely due to the object itself and called self-screening. The same criterion applies to blanket screening when the Newtonian potential is essentially dominated by the environment and is large enough to reduce the scalar charge below $\beta_\infty$.  In the following, we will focus on objects which are self- screened only and are embedded in the cosmological background. This case is easier to analyse and maximises the potential variation of constants between these objects and terrestrial values.

At the non-linear level the previous models fall within the category of scalar-tensor theories defined by the Lagrangian
\be
S=\int d^4 x \sqrt{-g}(\frac{R}{16\pi G_N} -\frac{(\partial \phi)^2}{2} -V(\phi))+S_m (\psi, A^2(\phi) g_{\mu\nu})
\label{act}
\ee
where $A(\phi)$ is an arbitrary function. The coupling to matter that we
have already introduced is simply given by
\be
\beta (\phi)= m_{\rm Pl} \frac{d \ln A(\phi)}{d \phi}.
\ee
The most important feature of these models  is that the scalar field dynamics are determined by an effective potential which takes into account the presence of the conserved matter density $\rho$ of the environment
\be
V_{\rm eff}(\phi) =V(\phi) +(A(\phi)-1) \rho.
\ee
With a decreasing $V(\phi)$ and an increasing $A(\phi)$, the effective potential acquires a matter dependent minimum
$\phi_{}(\rho)$ where the mass is also matter dependent $m(\rho)$.
Scalar-tensor theories whose effective potential $V_{\rm eff}(\phi)$ admits a density dependent minimum $\phi (\rho)$ can all be described
parametrically from the sole knowledge of the mass function $m(\rho)$ and the coupling $\beta (\rho)$ at the minimum of the potential \cite{Brax:2012gr,Brax:2011aw}
\be
\frac{\phi (\rho)-\phi_c}{m_{\rm Pl}}= \frac{1}{m_{\rm Pl}^2}\int_{\rho}^{\rho_c} d\rho \frac{\beta (\rho) A(\rho)}{m^2(\rho)},
\ee
where we have identified the mass as the second derivative
\be
m^2 (\rho)= \frac{d^2 V_{\rm eff}}{d\phi^2}\vert_{\phi=\phi (\rho)}
\ee
and the coupling
\be
\beta (\rho)= \frac{d\ln A}{d\phi}\vert_{\phi=\phi(\rho)}.
\ee
For most models, in the appropriate density range from a few $g/{\rm cm}^3$ to cosmological densities, the function $A(\rho)$ is essentially constant and equal to one. This is a phenomenological requirement which follows from the fact that the Einstein frame mass of a fermion is given by $A(\rho) m_\psi^{(0)}$ where $m_\psi^{(0)}$ is the bare mass in the matter Lagrangian. The Einstein frame masses must be almost constant throughout the cosmological history from Big Bang Nucleosynthesis (BBN) (where the density is similar to a few $g/{\rm cm}^3$'s) to the critical density of the Universe now.
In the following, we shall only consider models where $A(\rho)\sim 1$, $m(\rho)$ increases with $\rho$ as befitting the chameleon mechanism and $\beta (\rho)$ decreases with $\rho$ increasing as befitting the
Damour-Polyakov mechanism. This implies that $\phi(\rho)$ is a decreasing function of $\rho$.

It is often simpler to characterise the functions $m(\rho)$ and $\beta(\rho)$ using the time evolution of the matter density of the Universe
\be
\rho(a)=\frac{\rho_0}{a^3}
\ee
where $a$ is the scale factor whose value now is $a_0=1$. This allows one to describe characteristic models in a simple way.

\subsection{f(R) and dilaton}

First, a large class of interesting models  with a screening mechanism of the chameleon type consists of the large curvature $f(R)$ models \cite{Hu:2007nk} corresponding to the action
\be
S=\int d^4x \sqrt{-g} \frac{f(R)}{16\pi G_N}
\ee
where the function $f(R)$ is expanded in the large curvature regime
\be
f(R)=\Lambda + R -\frac{f_{R_0}}{n} \frac{R_0^{n+1}}{R^n}
\ee
where $\Lambda$ is a cosmological constant term and $R_0$ is the present day curvature.  These models can be reconstructed using the constant $\beta(a)=1/\sqrt{6}$ and the mass function as a function of the scale factor $a\le 1$
\be
m(a)= m_0 (\frac{4\Omega_{\Lambda 0}+ \Omega_{m0} a^{-3}}{4\Omega_{\Lambda 0}+ \Omega_{m0}})^{(n+2)/2}
\ee
where the mass on large cosmological scale is given by
\be
m_0= H_0 \sqrt{\frac{4\Omega_{\Lambda 0}+ \Omega_{m0} }{(n+1) f_{R_0}}},
\ee
$\Omega_{\Lambda 0} \approx 0.73$, $\Omega_{m0}\approx 0.27$ are the dark energy and matter density fractions now \cite{Brax:2012gr}. Local tests of gravity require that (see later)
\be
f_{R_0}\lesssim 0\cdot 66 \ 10^{-7}
\ee
which will be our benchmark value in the following.

Another relevant example is the environmentally dependent dilaton \cite{Brax:2010gi} where the screening mechanism is of the Damour-Polyakov type. This model is inspired by string theory in the large string coupling limit
with an exponentially runaway potential
\be
V(\phi)=V_0 e^{-\frac{\phi}{m_{\rm Pl}}}
\ee
where $V_0$ is determined to generate the acceleration of the Universe now and the coupling function is
\be
A(\phi)=\frac{A_2}{2m_{\rm Pl}^2} (\phi-\phi_\star)^2.
\ee
In dense environments, the coupling to matter vanishes as $\phi\to\phi_\star$. The coefficient $A_2$ is chosen in such a way that local tests of gravity are satisfied (see later).
These models can be described using the coupling function
\be
\beta(a)= \beta_0 a^3
\ee
where $\beta_0$ is related to $V_0$ and is determined by requiring that $\phi$ plays the role of late time dark energy which sets $\beta_0=\frac{\Omega_{\Lambda 0}}{\Omega_{m0}}\sim 2.7$, and the mass function which reads
\be
m^2(a)= 3 A_2 H^2(a)
\ee
and is proportional to the Hubble rate with the mass on cosmological scales now given by $m_0=\sqrt{3 A_2} H_0$.
\begin{figure}
\centering
\includegraphics[width=0.50\linewidth]{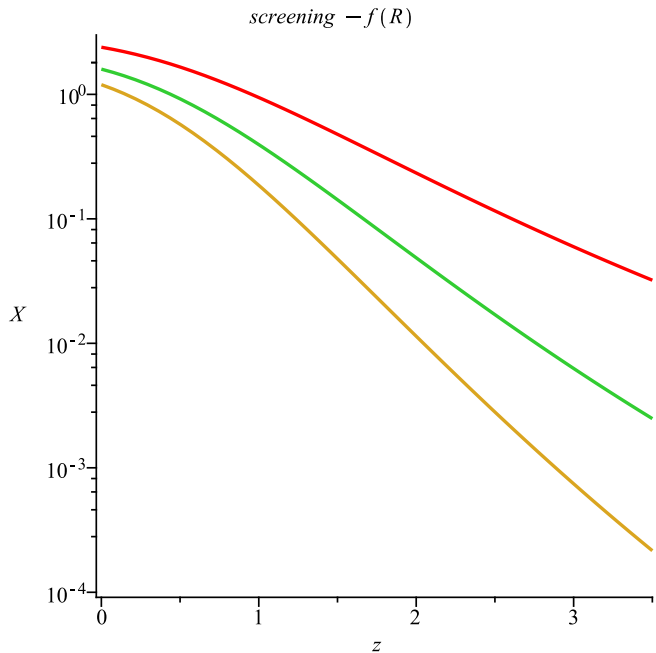}
\caption{The screening criterion   $\Phi_N\lesssim \frac{H_0^2}{m_0^2} X(z)$ where $X(z)$ is represented here for $f(R)$ models with $n=1,2,3$ from top to bottom for astrophysical objects of the galactic type with $a_c\sim a_G$. Unscreened objects are such $\frac{\Phi_N m_0^2}{H_0^2}$ falls below the curves for each $n$. Local gravitational tests imply that $H_0/m_0\lesssim 10^{-3}$ for $f(R)$ models, implying that unscreened objects must have a Newtonian potential less than $10^{-6}$ at low redshift and smaller at higher redshift.}
\end{figure}

\subsection{Screening}

The tomographic map allows one to reformulate the screening criterion as
\be
\frac{1}{m_{\rm Pl}^2}\int_{\rho_\infty}^{\rho_c} d\rho \frac{\beta (\rho)}{m^2(\rho)}\lesssim  2\beta_\infty \Phi_N
\ee
where $\rho_\infty$ is the density far away from the object.
When this inequality is not satisfied, the object is unscreened. The unscreening condition for an astrophysical object embedded in the cosmological background at redshift $z$ can be expressed using the relation
between $\rho$ and $a(z)=1/(1+z)$. This necessary condition for  astrophysical objects  to be  unscreened  is that their Newtonian
potential satisfies
\be \Phi_N \lesssim \frac{H_0^2}{m_0^2} X(z) \ee
where we have introduced the function
\be
X(z)= \frac{9}{2f(a(z))}\int_{a_{c}}^{(z+1)^{-1}} \frac{f (a) \Omega_m (a) H^2(a)}{a g(a)}  da
\ee
 and we have defined the dimensionless functions $m^2(a)=m_0^2 g(a)$ and $\beta(a)= \beta_0 f(a)$. For the Milky Way we have $a_c\equiv a_G\sim 10^{-2}$. In this case, we have plotted the variation of $X(z)$ as a function of the redshift $z$ for $f(R)$ models and the dilaton in Fig.1 and Fig.2.
For low redshift objects, unscreened objects are characterised by
\be \Phi_N \lesssim \frac{H_0^2}{m_0^2}. \ee
This has a stringent consequence for the modified gravity models. Indeed, the Milky Way, which must be screened to avoid too much disruption in the dynamics of satellite galaxies \cite{Kesden:2006vz}, is such that  $\Phi_G\sim 10^{-6}$. The mass of the scalar field in the
cosmological background now $m_0$ must then satisfy \cite{Brax:2011aw,Wang:2012kj}
\be
\frac{m_0}{H_0} \gtrsim 10^3.
\ee
This implies that the cosmological range of the scalar field must be less than a few Mpc's now.
As a result we see that unscreened astrophysical objects must necessarily have a low Newtonian potential
\be
\Phi_N \lesssim 10^{-6}
\label{weak}
\ee
when they are at low redshift, and even smaller when their redshift is $z\gtrsim 1$ as $X(z)$ drops below a few percent for $f(R)$ models.

The screening of the earth, which is necessary to evade local tests of gravity, implies that we have that
\be
\frac{\Delta \phi}{m_{\rm Pl}} \lesssim  2Q_\oplus \Phi_\oplus
\label{earth}
\ee
between the earth and any other location in the Milky Way. This follows from the fact that $\Delta \phi_\oplus= \phi_G-\phi_\oplus= 2 Q_\oplus \Phi_\oplus$ and $\Delta\phi\equiv \phi_c -\phi_\oplus \le \Delta\phi_\oplus$ as $\phi_c \le \phi_G$ as the density of any astrophysical system in the Milky Way is larger than the halo density.

This is an  important bound as it restricts the variation of $\phi$ extremely tightly. Indeed,
local tests of the strong equivalence principle in the solar system carried out by the Lunar Ranging experiment \cite{Williams:2012nc} imply that \cite{KhouryWeltman}
\be
Q_\oplus \le 10^{-7}
\ee
and the Newtonian potential on earth is $\Phi_\oplus \sim 10^{-9}$.
This gives strong constraints on the models as it fixes the values of
\be
\frac{9}{2}\int_{a_{\oplus}}^{a_G} \frac{\beta^ (a) \Omega_m (a) H^2(a)}{a m^2(a)}  da= Q_\oplus \Phi_\oplus \sim 10^{-16}.
\ee
For instance for large curvature $f(R)$ models,  this is a weaker condition than the screening of the Milky Way when $n\gtrsim 1$. For dilaton models on the other hand, this is a stronger condition than the screening of the Milky Way and requires that $A_2 \gtrsim 4\cdot 10^9$.

Effects on the variation of constants will  appear to be maximal for unscreened objects. This selects only a particular type of astrophysical absorbers as potential candidates.
For instance, dwarf galaxies with Newtonian potentials of order $\Phi_N={\cal{O}}(10^{-8}-10^{-7})$  \cite{Vikram:2013uba} satisfy the criterion (\ref{weak}) and could be unscreened for $f(R)$ models. On the other hand, these galactic objects are screened for dilaton models with $A_2 \gtrsim 4\cdot 10^9$. Similarly intergalactic  Lyman limit systems with densities as low $n\sim 10^{-4}\ \rm{cm}^{-3}$ corresponding to $a_L\sim 0.21$ and sizes $R={\cal O}(1)\ {\rm kpc}$  such that
$\Phi_N={\cal{O}}(10^{-13})$ could also be unscreened for both $f(R)$ and dilaton models. We will come back to the types of variations of constants that one could envisage for these objects in the discussion section.

Importantly,
notice that the screening criterion depends on Newton's potential. It also depends on the density of the object in a mild manner via $a_c$.
We have represented in Fig. 1  the function $X(z)$ for galactic absorbers and $f(R)$ models. The unscreening criterion becomes more difficult to fulfil as $z$ increases as $X(z)$ decreases to low values. We have also compared the function $X(z)$ for different values of $a_c=a_G$ and $a_c=a_L$ in Fig. 3. The dependence of the object's density is only relevant at high enough redshift and negligible at low redshift.
We will analyse the variation of constants in this context in the following section.
\begin{figure}
\centering
\includegraphics[width=0.50\linewidth]{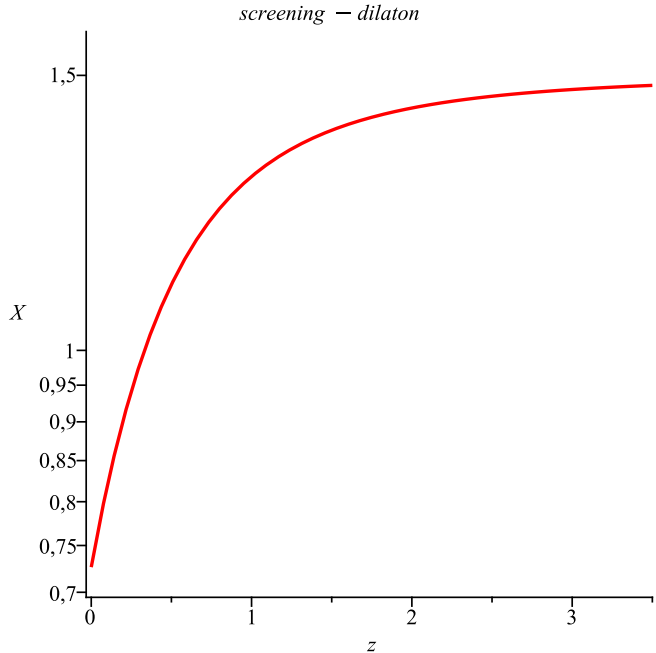}
\caption{The screening criterion $\Phi_N \lesssim \frac{H_0^2}{m_0^2} X(z)$ where $X(z)$ is represented here for dilaton models for astrophysical object of the galactic type with $a_c\sim a_G$. Unscreened objects are such $\frac{\Phi_N m_0^2}{H_0^2}$ is below the curve. Local gravitational tests imply that $H_0/m_0\lesssim 10^{-4}$ for dilaton models, so that unscreened objects must have a Newtonian potential less than $10^{-8}$.}
\end{figure}

\begin{figure}
\centering
\includegraphics[width=0.50\linewidth]{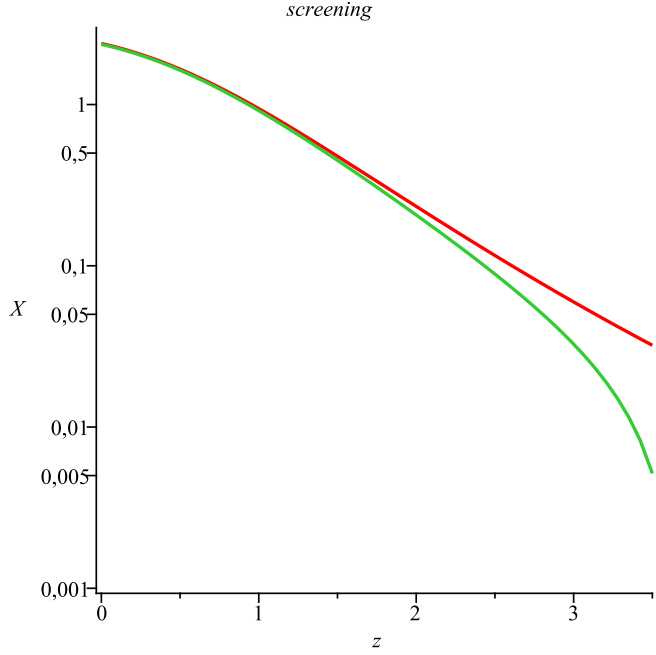}
\caption{The screening criterion $\Phi_N \lesssim \frac{H_0^2}{m_0^2} X(z)$ where $X(z)$ is represented here for $f(R)$  models with $n=1$ for two types of astrophysical objects with $a_c=a_G$ (top) and $a_c=a_L$ (bottom). No significant discrepancy appears unless the objects are at relatively high redshift.  }
\end{figure}

\section{Variation of Constants}

\subsection{Variation of $\mu$}

From this we can study the environmental dependence of both particle masses and the fine structure constant. In an environment where galaxies and gas clouds up to a redshift of a few can be screened or unscreened,
the scalar field develops a profile where it takes values equal to the density dependent minimum $\phi(\rho)$ in screened regions and the cosmological background value in unscreened parts of the Universe.
Hence there is no one to one dependence of fundamental constants on the redshift (or time) as in dark energy models, but an environment dependence which reflects how screened and unscreened regions are distributed in the Universe.
The variation of masses  follows from the relationship between the physical mass $m_\psi$ of fermions in the Einstein frame and the bare mass $m_\psi^{(0)}$ as
it would appear in the Lagrangian
\be
m_\psi= A(\phi) m_\psi^{(0)}
\ee
implying that the spatial variation of  masses is
\be
\frac{\Delta m_\psi}{m_\psi}=\Delta A
\ee
corresponding to the variation of $A$ between two locations in the Universe and we have use the fact that $A\sim 1$.
In particular, the proton to electron mass ratio varies like
\be
\frac{\Delta \mu}{\mu}=-\Delta A
\ee
where we have used the fact that the proton mass is dominated by the QCD scale which is scalar independent due to the conformal invariance of the gluon Lagrangian.
The tomographic map allows one to express the variation of $\mu$ between terrestrial values and far away ones as
\be
\frac{\Delta \mu}{\mu}=-9\int_{a_{\oplus}}^{a_{\rm abs}} \frac{\beta^2 (a) \Omega_m (a) H^2(a)}{a m^2(a)}  da
\ee
where $a_\oplus\sim 10^{-8}$ corresponds to the density in the atmosphere where particle masses are known and $a_{\rm abs}$ is either the redshift such that $\rho(a_{\rm abs})$ is the density of the far away screened region where the absorber lies or $a_{\rm abs}=a(z)$ if the absorbers are in an unscreened part of the Universe at a redshift $z$ with $a(z)=1/(1+z)$.

For theories like $f(R)$ where $\beta(a)=\beta$ is a constant we have a simplified expression
\be
\frac{\Delta \mu}{\mu}=-\beta \frac{\Delta \phi}{m_{\rm Pl}}.
\ee
In general, the variation of $\beta$ as a function of the density needs to be taken into account and the variation of $\mu$ is not directly proportional to the variation of $\phi$.
For unscreened objects located at redshift $z$, the variation of $\mu$ is given by
\be
\frac{\Delta \mu}{\mu}=-9\int_{a_{\oplus}}^{(z+1)^{-1}} \frac{\beta^2 (a) \Omega_m (a) H^2(a)}{a m^2(a)}  da.
\ee
Notice that this is very different from the usual variation of $\mu$ for dark energy models where the field $\phi(z)$ rolls and evolves with $z$. Here, only unscreened objects are sensitive to the
background cosmological evolution of the scalar field.
This result is also different from the BSBM theory where the variation of constants is proportional to the variation of Newton's potential. In fact the screening effect on the variation of $\mu$ can be better understood for theories where $\beta(a)$ increases with $a$, i.e. the coupling is weaker  in dense environments
\be
\vert \frac{\Delta \mu}{\mu}\vert  \le \beta_0 \frac{\Delta \phi}{m_{\rm Pl}}
\label{bp}
\ee
which expresses the fact that the case with a constant $\beta (a)=\beta_0$ leads a maximal variation. Using the fact that $\phi(\rho)$ increases when $\rho$ decreases, we have for unscreened regions $\phi(a_{\rm abs})-\phi_\oplus= (\phi(a_{\rm abs})-\phi_G) +(\phi_G -\phi_\oplus)$ together with $\phi_G -\phi_\oplus=2Q_\oplus \Phi_\oplus$ where $\phi_G$ is the field value in the Milky Way and
$\phi(a_{\rm abs})-\phi_G\le \phi_0-\phi_G\le 2 \beta_0 \Phi_G$ because of the screening of the Milky Way. As a result we have the upper bound
\be
\vert \frac{\Delta \mu}{\mu}\vert  \le 2\beta_0^2 \Phi_G.
\label{up}
\ee
This bound is model independent and only results from the screening of the Milky Way and the earth with $Q_\oplus \Phi_\oplus \ll \beta_0 \Phi_G$.
Therefore we find the upper bound on the variation of $\mu$
\be
\vert \frac{\Delta \mu}{\mu}\vert \lesssim 10^{-6}
\ee
when $\beta_0 = {\cal O}(1)$.

For screened objects similar to the Milky Way, where $a_{\rm abs}\sim a_G$, the screening effect reduces drastically the magnitude of the variation of $\mu$. Indeed, let us first focus on
probes in the local galactic environment.  Using (\ref{bp}), the bound on $\Delta\phi$ (\ref{earth}) coming from the screening of the earth implies that
\be
\vert \frac{\Delta \mu}{\mu}\vert  \lesssim 2\beta_0  Q_\oplus \Phi_\oplus \lesssim 10^{-16} \beta_0.
\ee
For models with $ \beta_0={\cal O}(1)$ such as $f(R)$ and the dilaton, this is 8 orders of magnitude lower than the present experimental bound \cite{Levshakov:2013oja} in the Milky Way.
More generally, this bound applies to all galactic environments similar to the one of the Milky Way as $a_{\rm abs}\sim a_G \sim 10^{-2}$ and can only be evaded in potentially unscreened regions of the Universe such as molecular clouds in dwarf galaxies.

We have shown in Fig. 4 the variation of the proton to electron mass ratio deduced from transitions occurring in distant unscreened objects at redshift $z$
for large curvature $f(R)$ models with $\vert f_{R0}\vert =0.66\ 10^{-6}$ and $n=1,2,3$. The variation is maximal $\frac{\Delta \mu}{\mu}\sim 3.3\cdot 10^{-7}$ at $z\sim 0$ and drops to $10^{-9}$ for $z\sim 3$.
In Fig.5 we have the same variation for a dilaton model with $A_2=4\cdot 10^9$. The maximal variation is much smaller than  in the $f(R)$ case. This follows from the fact that dilaton models are constrained by local experiments in
a much more stringent way than $f(R)$ models. This reduces the allowed range of the cosmological interaction mediated by the dilaton and implies that the field varies less spatially.

Notice that the $f(R)$ models saturate the upper bound (\ref{up}) for unscreeened objects at small redshift and $\beta_0=1/\sqrt{6}$. Moreover, such a variation of $\mu$ as shown in Fig. 4 is within the reach of future observations for unscreened objects such as dwarf galaxies. In this sense, $f(R)$ models are  optimal to look for a variation of constants in modified gravity.

\begin{figure}
\centering
\includegraphics[width=0.50\linewidth]{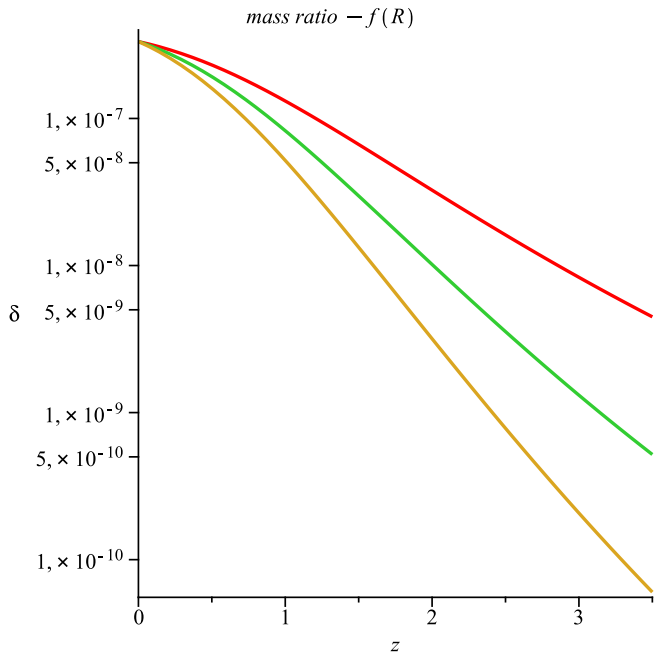}
\caption{The variation of the  proton to electron mass ratio $\delta=\vert \frac{\Delta \mu}{\mu}\vert$ for unscreened absorbers with $a_c=a_G$, e.g.  molecular clouds of dwarf galaxies,  as a function of the redshift of the absorbing system. The three curves represent the case of large curvature $f(R)$ models, when $\vert f_{R0}\vert =0.66\cdot 10^{-6}$,  with $n=1,2$ and $3$ respectively from top to bottom.}

\label{hole}
\end{figure}

\subsection{Variation of $\alpha$}

We now turn to the spatial variation of the fine structure constant. For dark energy models with a very low mass scalar field, this was investigated in \cite{Carroll:1998zi} where the coupling of the scalar field to electromagnetism was taken to be linear. In modified gravity models defined here and at the classical level, the electromagnetic Lagrangian (\ref{act}) is independent of $\phi$. This follows from the conformal invariance of the gauge kinetic terms. A scalar dependence of the fine structure constant appears due to quantum effects only. There are two origins to this coupling.
First,  the scalar field couples to the fermion kinetic terms according to $A^3(\phi)\bar\psi_E\gamma^\mu D_\mu \psi_E$. The fermions can be normalised and such a coupling effaced provided one performs
a scalar dependent field redefinition of $\psi=A^{3/2}(\phi)\psi_E$. This rescaling induces a coupling to photons which can be seen as the result of the one loop diagram involving 2 photons, one scalar and the fermions of the standard models running in the loop. These
quantum effects are captured by the change of the fermion measure in the path integral of  modified gravity theories \cite{Brax:2010uq} and can be calculated using the Fujikawa method. For the normalised fermions, the mass term of each fermion species is scalar dependent $ A(\phi) m_\psi^{(0)}\bar\psi\psi$. At low energy, when the massive fermions decouple and are integrated out, the triangle diagrams involving 2 photons, a scalar mass insertion and the fermions running in the loop lead to another contribution to the scalar dependence of the fine structure constant \cite{Brax:2009ey}. The two  effects  imply that there is a coupling of the scalar field to photons
\be
\beta_\gamma (\phi)= \frac{5\alpha_0 N_f}{6\pi} \beta(\phi)
\ee
where $\alpha_0$ is the fine structure constant as it appears in the QED Lagrangian without any dependence on the scalar field $\phi$, and this relation stands for any background value $\phi$. The number of fermions $N_f$ can be taken to be the one in the standard model $N_f= 12$ where
there are 6 quark families and 6 lepton families \footnote{We have integrated out the 3 neutrino families as we are interested in low energy physics at the dark energy scale. If some of the neutrinos have a mass below this scale, only the numerical factor changes as can be extracted from \cite{Brax:2010uq}}. The effective fine structure constant becomes now
\be
\frac{1}{\alpha}= \frac{A_\gamma (\phi)}{\alpha_0}
\ee
where we have
\be
A_\gamma (\phi)= (A(\phi))^\eta
\ee
with $\eta=\frac{5\alpha_0 N_f}{6\pi}$.
This implies that the fine structure has a spatial variation related to the environment dependence of the proton to electron mass ratio
\be
\frac{\Delta \alpha}{\alpha}=- \frac{5\alpha_0 N_f}{6\pi}\frac{\Delta \mu}{\mu}.
\ee
We expect the spatial variation of the fine structure constant to be at least one order of magnitude lower than the proton to electron mass ratio and at most of order $10^{-7}$. We have represented in Fig. 6 and Fig. 7 the variation of $\alpha$ for unscreened absorbers similar to Lyman limit systems with $a_c=a_L$. For $f(R)$ models, the maximal deviation is of the order of $10^{-8}$ and lower for dilaton models.

\begin{figure}
\centering
\includegraphics[width=0.50\linewidth]{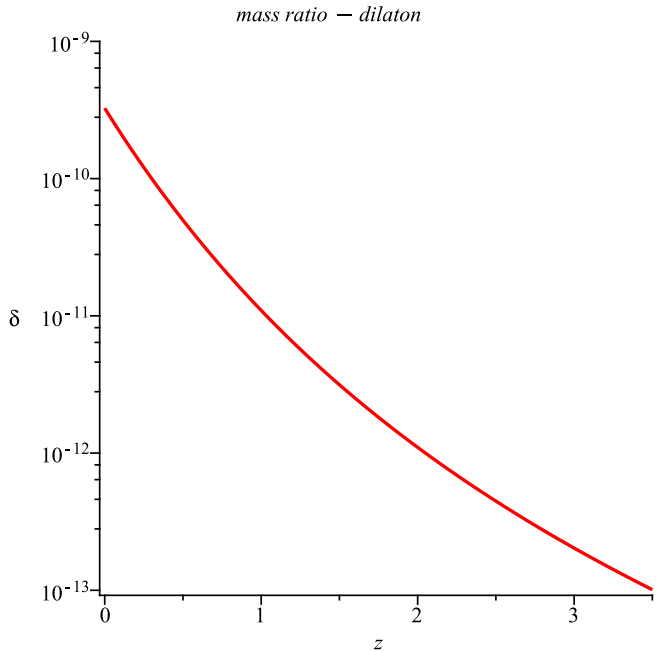}
\caption{The variation of the  proton to electron mass ratio $\delta=\vert \frac{\Delta \mu}{\mu}\vert$ for unscreened absorbers with $a_c=a_G$, e.g. molecular clouds of dwarf galaxies,  as a function of the redshift of the absorbing system for a dilaton model with $\beta_0=2.7$ and $A_2=4\cdot 10^9$.}

\end{figure}

\begin{figure}
\centering
\includegraphics[width=0.50\linewidth]{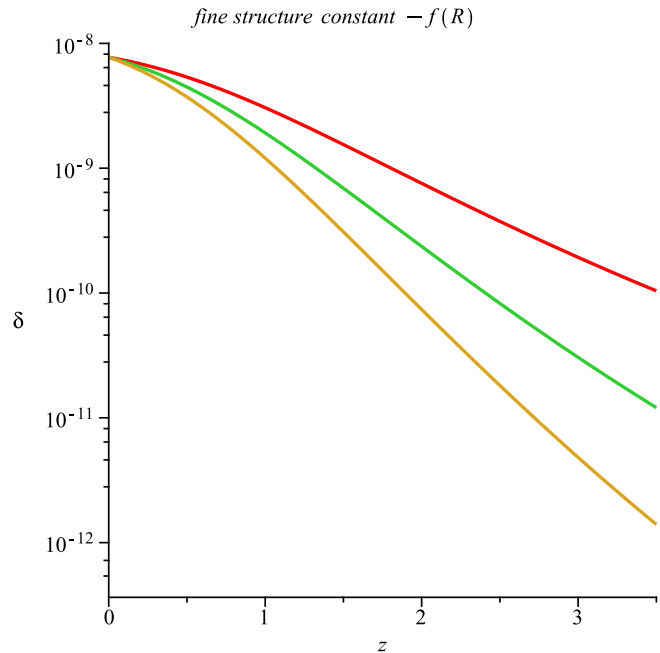}
\caption{The variation of the fine structure constant for unscreened absorbers like Lyman limit systems with $a_c=a_L$  as a function of the redshift of the absorbing system. The three curves represent the case of large curvature $f(R)$ models, when $\vert f_{R0}\vert =0.66\cdot 10^{-6}$,  with $n=1,2$ and $3$ respectively from top to bottom.}

\label{hole}
\end{figure}

\begin{figure}
\centering
\includegraphics[width=0.50\linewidth]{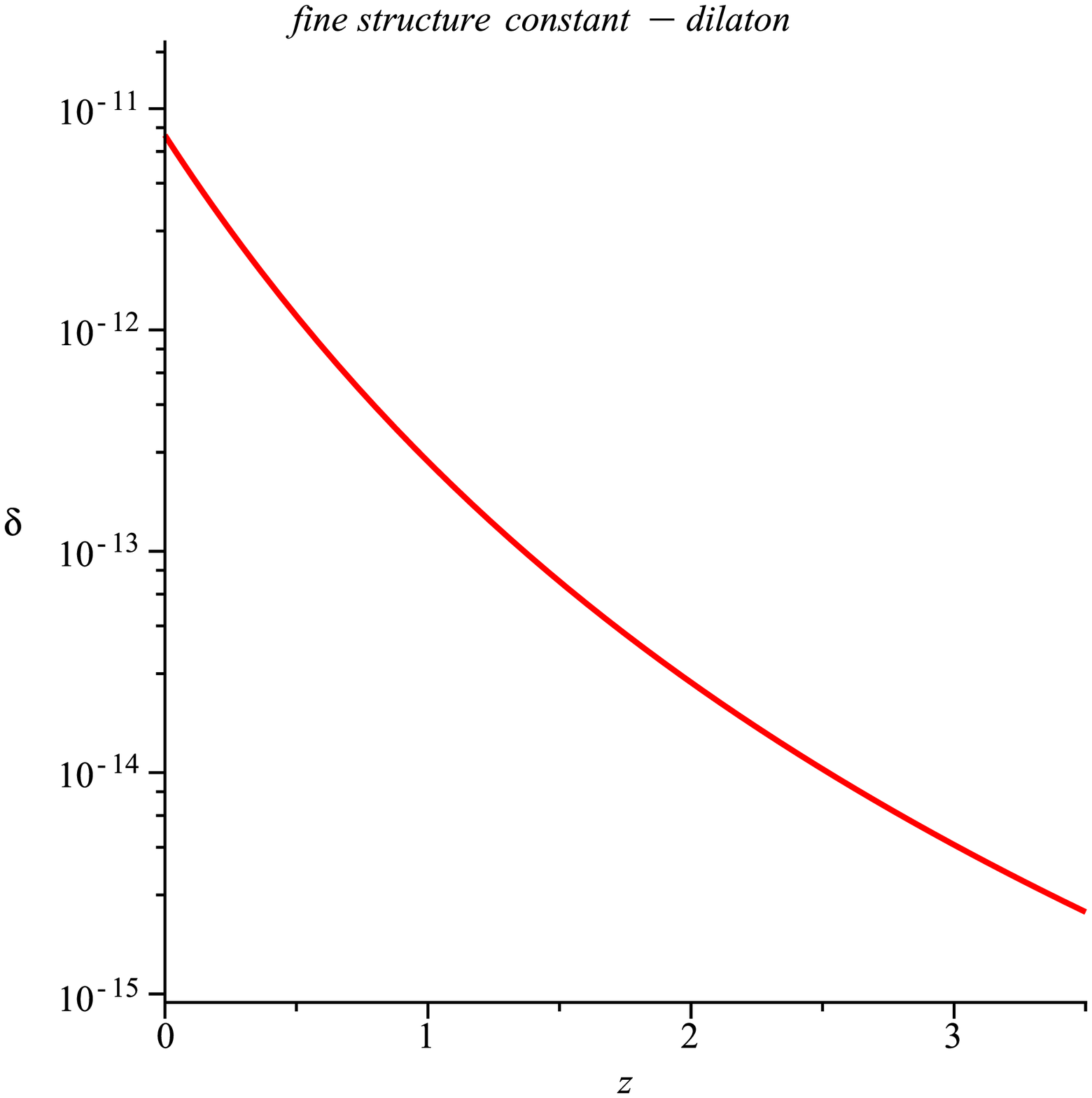}
\caption{The variation of the fine structure constant for unscreened absorbers like Lyman limit systems with $a_c=a_L$  as a function of the redshift of the absorbing system for a dilaton model with $A_2=4\cdot 10^9$.}

\label{hole}
\end{figure}

\section{Discussion}

\subsection{Current bounds}

Let us now compare our results to astrophysical observations \cite{Bonifacio:2013vfa}. In the Milky Way, the proton to electron mass ratio is well constrained by observing transitions lines in molecular clouds \cite{Molaro2008,Molaro:2009jc,Levshakov:2010cw,Levshakov:2010tj}. The best bound in the Milky Way is
\be
\frac{\Delta \mu}{\mu}< 2\times 10^{-8}
\ee
at the 3-$\sigma$ level \cite{Levshakov:2013oja,Levshakov:2013ufa}. As already mentioned, the tests of the equivalence principle with the earth-moon system imply that screened models in environments similar to the Milky Way cannot lead to a variation larger than $10^{-16}$, hence such galactic observations of molecular clouds to detect a variation of $\mu$ cannot probe screened models. For redshifts $z\le 1$, the best bound is
\be
\frac{\Delta \mu}{\mu}<(0\pm 1)\times 10^{-7}
\ee
at $z=0.89$ \cite{Bagdonaite:2013sia}. Again this bound is obtained from methanol transitions in a galactic environment similar to the Milky Way where the screening effect implies that any variation of $\mu$ for screened modified gravity is much lower. Observations of transition lines at larger redshifts are obtained for instance using Damped Lyman-$\alpha$ systems where the column density is comparable to the one in the Milky Way ${\rm N(H1)}\gtrsim 10^{20}\ {\rm cm}^{-2}$. In this case, the best observational bound is \cite{Wendt:2013zna,Wendt:2013nca}
\be
\frac{\Delta \mu }{\mu} <10^{-5} ,
\ee
much larger than the expected $10^{-16}$ level for screened models in such galactic environments.
There are claims that a non-zero deviation of $\frac{\Delta \mu}{\mu}= (12.7\pm 4.5_{\rm stat}\pm 4.2_{\rm sys})\times 10^{-6}$ has been observed \cite{Bagdonaite:2013eia}  but systematic effects may not have been completely taken into account \cite{Wendt:2013zna}. If confirmed, such a deviation would invalidate the screened models.
The observations of the variation of $\alpha$ seem to indicate that $\alpha$ could have a dipolar variation at the $10^{-5}$ level \cite{King:2012id}. This is not confirmed  by more recent analyses \cite{Molaro:2013saa}. If the dipolar variation of $\alpha$ were to be confirmed this would rule out the screened modified gravity models presented in this paper.
Finally let us consider the Earth-bound atomic clock experiments which constrain the proportionality factor $k_\alpha$ between the spatial variation of the fine structure constant and the gravitational potential, here due to the
seasonal variation of the Newtonian potential of the sun evaluated at the earth
\be
\frac{\delta\alpha}{\alpha}=k_\alpha \delta \Phi_N.
\ee
 In the screened models we have
\be k_\alpha= \frac{5\alpha_0 N_f}{3\pi} \beta_\oplus Q_\oplus \frac{\Phi_\oplus}{\Phi_\odot}\ee
where the ratio between the Earth and the sun Newtonian potentials is $\frac{\Phi_\oplus}{\Phi_\odot}={\cal O}(10^{-3})$, implying that $k_\alpha \lesssim 10^{-11}$, which is four orders of magnitude lower than the experimental bound \cite{Leefer:2013waa}.

\subsection{Prospects}

We will outline some of the consequences of our results for future observations. This discussion will be far from exhaustive and would require a more thorough investigation which is far beyond the scope of this paper. Here we limit ourselves to some indications about which models could potentially be tested with the variation of constants and which absorbing systems could possibly be of relevance. A more thorough study is left for future work.

We focus exclusively on $f(R)$ models in the large curvature regime as a guideline as they potentially give larger deviations of $\mu$ and $\alpha$ than models like the dilatons where the coupling to matter decreases fast with the matter density. Moreover, as we have already seen, these model saturate the bound (\ref{bp}) for low redshift astrophysical objects which are unscreened when the parameter $f_{R_0}=0.66 \ 10^{-7}$ is the largest possible value
which still allows for a screening of the Milky Way. This bound on $f_{R_0}$ is stronger than any present cosmological bound \cite{Lombriser:2014dua}. A more stringent  bound $f_{R_0}\lesssim 5 \cdot 10^{-7}$ has been obtained using distance indicators from unscreened  dwarf galaxies in the local environment \cite{Jain:2012tn} and more recently by comparing the dynamics of the stellar and gaseous components of dwarf galaxies \cite{Vikram:2013uba} implying that $f_{R_0}\lesssim 10^{-7}$.

The bound (\ref{weak}) implies that only astrophysical objects with low surface Newtonian potentials can be unscreened and lead to a substantial variation of $\mu$. The first type of object one may envisage comprises
the dwarf galaxies with $\Phi_N$ ranging from $10^{-8}$ to $10^{-7}$. In this case, the screening criterion is satisfied for dwarf galaxies at low redshift and the variation of $\mu$ could be tested using molecular clouds.
At low redshift, and using $f_{R_0}\lesssim 10^{-7}$, we expect a variation of $\mu$ to be of the order of $10^{-8}$. This is within the range of future observations. If no effect were to be observed, this would lower the admissible values of $f_{R_0}\lesssim 10^{-8}$.

Below this value, only astrophysical objects with very low Newtonian potentials could lead to a substantial variation of constants. One possibility would be to utilise extragalactic Lyman limit systems with $\Phi_N \sim 10^{-13}$.
Unfortunately, these objects cannot be used to test the variation of $\mu$ as they are not dense enough to prevent the photo-dissociation of molecules. On the other hand, they are candidates for a variation of $\alpha$.
These intergalactic clouds with low hydrogen column densities
can be observed in resonance lines of atoms and ions, and a variation of $\alpha$ could be inferred, at best at the $10^{-10}$ level for $f(R)$ models. If this extreme level of precision could be achieved, these would be the most sensitive tests of the $f(R)$ models, used as templates of modified gravity. Unfortunately, the expected sensitivity for near future observations is at best at the $10^{-8}$ for a variation of $\alpha$, postponing the possibility of testing very low values of $f_{R_0}$ to the more distant future \cite{Pro}.

\section{Conclusion}

The spatial dependence of fundamental constants due to the environment in screened modified gravity is very different from the usual time or redshift dependence in dark energy models \cite{Thompson:2012pja}. Here we have shown that what matters
is not the cosmological dynamics of the field but the way it gets screened or unscreened in different regions of the Universe. As a result, two absorbing regions at the same redshift would show different values of the fundamental constants if they lie in screened and unscreened regions respectively. This could lead to new ways of analysing data where a tomographic description of the Universe, mapping screened and unscreened regions \cite{Cabre:2012tq}, would be correlated to the measured variations of constants. A strong correlation between the two maps would indicate that a modification of gravity is at play. Of course, for this, the precision of the observations should be high as we have shown that the fine structure constant is expected to vary at most at the $10^{-7}$ level and the proton to electron mass ratio at the $10^{-6}$ one. This should be within the reach of forthcoming experiments \cite{poj}, for instance  using molecular clouds in dwarf galaxies to test the variation of $\mu$.

\acknowledgments I would like to thank S. Levshakov for extremely simulating exchanges and P. Valageas for suggestions on the manuscript.

\end{document}